\documentclass[twocolumn,preprintnumbers,prl,amsmath,amssymb,superscriptaddress,showpacs]{revtex4}
\usepackage{epsfig}
\usepackage{graphicx}

\def\bra#1{\langle#1|} \def\ket#1{|#1\rangle}

\begin{document}

\title{Experimental Proof of Quantum Nonlocality without Squeezing}

\author{J. Niset}
\affiliation{Quantum Information and Communication, Ecole Polytechnique, CP 165, Universit\'e Libre de
Bruxelles, 1050 Brussels, Belgium}

\author{A. Acin}
\affiliation{ICFO-Institut de Ciencies Fotoniques, Mediterranean
Technology Park, 08860 Castelldefels (Barcelona), Spain}

\author{U. L. Andersen}
\affiliation{Institut f\"{u}r Optik, Information und Photonik,  Max-Planck Forschungsgruppe, Universit\"{a}t
Erlangen-N\"{u}rnberg, G\"{u}nther-Scharowsky str. 1, 91058, Erlangen, Germany}

\author{N. J. Cerf}
\affiliation{Quantum Information and Communication, Ecole Polytechnique, CP 165, Universit\'e Libre de
Bruxelles, 1050 Brussels, Belgium}

\author{R. Garc\'\i a-Patr\'on}
\affiliation{Quantum Information and Communication, Ecole Polytechnique, CP 165, Universit\'e Libre de
Bruxelles, 1050 Brussels, Belgium}

\author{M. Navascues}
\affiliation{ICFO-Institut de Ciencies Fotoniques, Mediterranean
Technology Park, 08860 Castelldefels (Barcelona), Spain}

\author{M. Sabuncu}
\affiliation{Institut f\"{u}r Optik, Information und Photonik,  Max-Planck Forschungsgruppe, Universit\"{a}t
Erlangen-N\"{u}rnberg, G\"{u}nther-Scharowsky str. 1, 91058, Erlangen, Germany}

\begin{abstract}
It is shown that the ensemble $\{p\,(\alpha),\vert\alpha\rangle\vert\alpha^*\rangle\}$ where
$p\,(\alpha)$ is a Gaussian distribution of finite variance and
$\vert \alpha \rangle$ is a coherent state can be better
discriminated with an entangled measurement than with any local strategy
supplemented by classical communication. Although this ensemble 
consists of products of quasi-classical states, it exhibits some quantum nonlocality. This remarkable effect is demonstrated experimentally by implementing the optimal local strategy together with a joint nonlocal strategy 
that yields a higher fidelity.
\end{abstract}


\maketitle 

Entanglement is known to be a valuable resource, which can be used
to achieve a large variety of quantum protocols, mostly through the
manipulation and detection of entangled states. Due to the 
nonlocal nature of entanglement, joint measurements are 
typically necessary to properly access the information stored in the quantum
correlations of entangled states. But entanglement has proven to
be much richer than that. It can, for example, take profit of purely
classical correlations to provide a better access to the information 
encoded in a product state \cite{Peres}. In particular, the optimal 
measurement of $N$ identically prepared qubits is 
known to be a joint entangled measurement \cite{Massar}. Moreover, 
some classical correlations perform better than others: a surprising 
result, obtained in \cite{Gisin}, is that more
information can be extracted from a pair of orthogonal qubits
than from two identical qubits.
An even more intriguing phenomenon, found in \cite{Bennett}, 
is the existence of ensembles
of orthogonal product states that cannot be perfectly distinguished 
using Local Operations and Classical Communications (LOCC) only, 
but can be perfectly discriminated through a joint, yet separable, 
measurement. This effect, named ``nonlocality without entanglement'',
suggests that quantum mechanics allows for
nonlocal behaviors that go much beyond entanglement \cite{fn}.

In the recent years, much attention has been devoted to quantum
information based on Continuous Variables (CV). Many results
initially derived for qubits have been successfully adapted to
infinite dimensional systems. It is thus tempting to ask 
whether one can observe nonlocal effects 
with CV product states. In this Letter, we answer this question
affirmatively by exhibiting an ensemble of bipartite product coherent states 
that can be better discriminated when the parties act globally 
rather than locally. Then, we report on the experimental demonstration 
of this nonlocal property based on the sideband encoding of a modulated laser 
beam with no need for squeezing. The superior discrimination of product states
via entangled measurements was verified experimentally only very recently
for two qubits, using a parametric down-conversion source and a linear-optics controlled-NOT gate \cite{Pryde}. To our knowledge, the present work
achieves the first experimental evidence of this form of nonlocality 
\cite{fn} with continuous variables.


We consider the ensemble of product states $\vert\alpha\rangle\vert\alpha^*\rangle$, with
$p\,(\alpha)$ being a Gaussian distribution, which 
was first introduced in the context of
the optimal CV phase-conjugation transformation \cite{Sofian}. 
Here, 
we show that this ensemble can be better discriminated if we have
access to joint operations rather than being restricted to LOCC.
In this respect, we exhibit an ensemble of quasi-classical states 
which are neither squeezed nor entangled, and nevertheless 
show some kind of nonlocality.
To estimate the quality of a particular measurement strategy we will use the mean fidelity, i.e.,
$$F=\sup_{M_y}\sup_{\rho_y}\sum_y\int  d\alpha
P(\alpha)\bra{\alpha}\bra{\alpha^*}M_y\ket{\alpha}\ket{\alpha^*}\bra{\alpha}\rho_y\ket{\alpha},$$
where $M_y$ are the positive operators defining the measurement,
$\sum_y M_y=1$, and $\rho_y$ are states prepared according to the
measurement results. This means that our goal is 
to optimally measure $\ket{\alpha}\ket{\alpha^*}$ and
prepare a state as close as possible to $\vert\alpha\rangle$. To
bound the fidelity achievable by local operations, we will first
show that the optimal LOCC strategies on
$\vert\alpha\rangle\vert\alpha^*\rangle$ and
$\vert\alpha\rangle\vert\alpha\rangle$ yield equal fidelities. We
next prove that the optimal measure-and-prepare strategy on identical
copies of $\vert\alpha\rangle$ is achieved by a local strategy. And finally, we
exhibit a nonlocal measurement on phase-conjugate states,
conjectured to be optimal, which gives a higher fidelity than the
optimal strategy on two identical copies. We will thus prove that
$$\label{relation2} F_L^*=F_L=F<F^*$$ where $F$ and $F^*$
($F_L$ and $F_L^*$) denote the optimal fidelities for global
(local) measurements on $\vert\alpha\rangle\vert\alpha\rangle$ and
$\vert\alpha\rangle\vert\alpha^*\rangle$ respectively.

Let us start by proving that local strategies on
$\vert\alpha\rangle\vert\alpha^*\rangle$ and
$\vert\alpha\rangle\vert\alpha\rangle$ give identical fidelities,
i.e. $F_L^*=F_L$. Recall that any LOCC strategy consists of a
sequence of correlated measurements plus a decision strategy
depending on the observed statistics. After $n$ rounds of
measurements, the relevant probabilities can be written as
\begin{equation}
Pr(\beta)=Tr\{(A_\beta \otimes B_\beta) \
\vert\alpha\rangle\langle\alpha\vert\otimes\vert\alpha^*\rangle\langle\alpha^*\vert\}
\end{equation}
with the positive operators $A_\beta$ and $B_\beta$ defined as
\begin{eqnarray*}
\nonumber A_\beta&=&A_{r_n}^n(r_1,r_2,\ldots,r_{n-1})\times\ldots\times  A_{r_3}^3(r_1,r_2)\times A_{r_1}^1\\
B_\beta&=&B_{r_{n-1}}^{n-1}(r_1,r_2,\ldots,r_{n-2})\times\ldots \times B_{r_2}^2(r_1)
\end{eqnarray*}
In this expression, $r_i$ is the outcome of the $i$th measurement,
and the upper index stands for its order in the sequence of measurements.
These operators depend on the decision strategy, and are
constrained by the measurement normalization conditions. Now,
suppose that a particular LOCC strategy is optimal for
$\vert\alpha\rangle\vert\alpha^*\rangle$ and gives the fidelity
$F_L^*$. We can easily map this optimal strategy into an optimal
LOCC strategy for $\vert\alpha\rangle\vert\alpha\rangle$. Indeed, using
\begin{equation*}
Tr\{A_\beta \vert\alpha\rangle\langle\alpha\vert \otimes
B_\beta\vert\alpha^*\rangle\langle\alpha^*\vert\}=Tr\{A_\beta \vert\alpha\rangle\langle\alpha\vert \otimes
B_\beta^*\vert\alpha\rangle\langle\alpha\vert\}
\end{equation*}
it is possible to define a LOCC sequence of measurements, only
replacing $B_\beta$ by $B^*_\beta$, achieving the same fidelity
$F_L^*$ for $\vert\alpha\rangle\vert\alpha\rangle$, hence
$F_L^*=F_L$. $\Box$

Next, let us prove that the optimal measure-and-prepare strategy
on $\vert\alpha\rangle\vert\alpha\rangle$ is a local strategy,
i.e. $F_L = F$.  Note that this result is already known for a
distribution of infinite width \cite{Stenholm}, using the
variances of the estimated quadratures as a figure of merit. Here,
we prove a more powerful result: we consider the realistic
case of finite distributions, and do not make any assumption on
the measurement nor the reconstruction. Actually, we prove
the more general result that the optimal strategy for $N$ copies
of a coherent state distributed according to a Gaussian of
variance $1/\lambda$ yields a maximum fidelity of
\begin{equation} \label{Fmax}
F^N\leq \frac{N+\lambda}{N+\lambda +1}
\end{equation}
which is exactly the fidelity achieved by $N$ independent heterodyne measurements and preparation of a coherent
state centered on $\frac{1}{N+\lambda} \sum_{i=1}^N \alpha_i$ (with $\alpha_i$ the result of the $i$th measurement).

Without loss of generality, we can restrict our
optimization to measurements consisting of projectors
$\vert\Phi_y\rangle\langle\Phi_y\vert $ and preparation of pure
states  $\vert\chi_y\rangle$. For the input states $\vert \alpha
\rangle^{\otimes N}$ distributed with the Gaussian distribution
$P(\alpha)=\frac{\lambda}{\pi}\exp(-\lambda\vert\alpha\vert^2)$,
the average fidelity reads
\begin{equation}
F=\sum_y\int  d\alpha \ P(\alpha)\vert\langle\alpha^{\otimes
N}\vert\Phi_y\rangle\vert^2\vert\langle\alpha\vert\chi_y\rangle\vert^2
.
\end{equation}
To bound this fidelity, we generalize a method used in
\cite{Hammerer} to calculate the optimal fidelity for a
measure-and-prepare strategy on a single copy of a coherent state
$\ket{\alpha}$  distributed according to $P(\alpha)$. First, one
needs to realize that we can concentrate the $N$ modes of $\vert
\alpha \rangle^{\otimes N}$ into one single mode $\vert
\sqrt{N}\alpha \rangle$ by mean of beam splitters. This operation
is unitary and completely reversible, hence will not change the
fidelity. We can thus write
\begin{eqnarray}\label{fidelity}
\nonumber F&=&\sum_y\int d\alpha P(\alpha)\vert\langle
\sqrt{N}\alpha\vert\phi_y\rangle\vert^2\vert\langle\alpha\vert\chi_y\rangle\vert^2\\
\nonumber &\leq&\sup_{\phi_y ,\,\chi_y}\sum_y\langle\chi_y\vert A_{\phi_y} \vert \chi_y\rangle\\
&=& \sup_{\phi_y}\sum_y\Vert A_{\phi_y} \Vert_\infty
\end{eqnarray}
after introduction of the operators
\begin{equation}
A_{\phi_y}=\int d\alpha
P(\alpha)\vert\langle\sqrt{N}\alpha\vert\phi_y\rangle\vert^2\vert\alpha\rangle\langle\alpha\vert
\end{equation}
The last equality of (\ref{fidelity}) is trivial as it is indeed
best to prepare the eigenstate of $A_{\phi_y}$ associated to the
largest eigenvalue for a given outcome $y$. We can now turn to the
core of the method. Following \cite{Hammerer}, we first prove that
\begin{equation} \label{expression1}
\Vert A_{\phi}\Vert_p \leq \frac{N+\lambda}{[(N+\lambda+1)^p-1]^{1/p}}\Vert A_\phi \Vert_1
\end{equation}
holds for all states $\vert \phi\rangle$ and all $p$ norms $\Vert
A\Vert_p=(Tr\{\vert A\vert ^p\})^{1/p}$. The limiting case
$p\rightarrow\infty$, in combination with the measurement
normalization $\sum_y\vert\phi_y\rangle\langle\phi_y\vert=1$
(which implies $\sum_y\Vert A_{\phi_y}\Vert_1=1$), is then
sufficient to prove equation (\ref{Fmax}). 
We only present here the main results of our calculation. An
interested reader should consult \cite{Hammerer} and
\cite{Shapiro} for more details.

The properties of the Trace allow us to write
\begin{eqnarray}\label{trace}
\nonumber \Vert A_\phi \Vert^p _p&=&Tr\{A_\phi ^p\}=Tr\{\vert\phi\rangle\langle\phi\vert^{\otimes p}B\}\\
\Vert A_\phi \Vert^p _1 &=& Tr\{A_\phi\}^p=Tr\{\vert\phi\rangle\langle\phi\vert^{\otimes p}C\}
\end{eqnarray}
where we have defined the operators $B$ and $C$ as
\begin{eqnarray*}
B&=&\iint d\alpha_1\ldots d\alpha_p P(\alpha_1)\ldots P(\alpha_p)\langle
\alpha_1\vert\alpha_2\rangle\ldots\langle \alpha_p\vert\alpha_1\rangle\\ &\times& \vert
\sqrt{N}\alpha_1\rangle\langle\sqrt{N}\alpha_1\vert\otimes\ldots\otimes\vert
\sqrt{N}\alpha_p\rangle\langle\sqrt{N}\alpha_p\vert\\ C&=&\bigotimes^p_{i=1}\int d\alpha_i P(\alpha_i)\vert
\sqrt{N}\alpha_i\rangle\langle\sqrt{N}\alpha_i\vert
\end{eqnarray*}
These two operators can be diagonalized in the same basis. A
unitary transformation turns them into tensor products of
unnormalized thermal states, which are diagonal in the
corresponding Fock state basis. Expressing the product state
$\vert \phi\rangle^{\otimes p}$ in this Fock state basis and
remembering that $Tr\{\vert\phi\rangle\langle\phi\vert^{\otimes
p}B\}\geq 0$, one finds that the two traces of (\ref{trace})
relate as
\begin{equation}
Tr\{\vert\phi\rangle\langle\phi\vert^{\otimes p}B\}\leq
\frac{(N+\lambda)^p}{(N+\lambda+1)^p-1}Tr\{\vert\phi\rangle\langle\phi\vert^{\otimes p}C\}
\end{equation}
The $p$-th root of this expression gives directly relation
(\ref{expression1}) and completes the proof. $\Box$

Now, let us prove the existence of a nonlocal
measurement on phase-conjugate states that yields a higher
fidelity than (\ref{Fmax}) for $N=2$. 
One such measurement was introduced in \cite{Sofian}
in order to show that the $\vert \alpha \rangle\vert\alpha^*\rangle$
encoding outperforms $\vert \alpha \rangle\vert\alpha\rangle$
in the case $\lambda=0$.
The strategy is the following; first the two modes
are sent on a Beam Splitter (BS), which outputs two coherent
states displaced along the $x$ and $p$ axis respectively, i.e.
$\vert \alpha \rangle\vert\alpha^*\rangle\rightarrow\vert
x_\alpha\rangle\vert p_\alpha\rangle $ (where $\alpha = (x_\alpha
+ i p_\alpha)/\sqrt{2}$). Next, the appropriate
quadratures are measured, 
and some state $\vert f_\beta\rangle$ is reconstructed according to the measurement outcomes.
In order to calculate the fidelity associated to this
strategy and easily compare with (\ref{Fmax}), we will suppose that we
have at our disposal $N$ coherent states, that is $N/2$ pair
$\vert \alpha \rangle\vert\alpha^*\rangle$, or equivalently one
pair $\vert \sqrt{N/2}\; \alpha
\rangle\vert\sqrt{N/2}\; \alpha^*\rangle$. The corresponding
fidelity reads
\begin{eqnarray}
\nonumber F_{BS}^{*N}&=&\iint P(\alpha)\ P\big(x,p \vert
x_\alpha,p_\alpha\big)\vert\langle
f_\beta\vert\alpha\rangle\vert^2 dx dp d\alpha\\
&=&2\frac{\lambda}{\pi^2}\int  e^{-2\vert\beta\vert^2}\langle
f_\beta\vert \hat{O}_\beta  \vert f_\beta\rangle d\beta
\end{eqnarray}
where $P\big(x,p \vert x_\alpha,p_\alpha\big)$ is the probability
to measure $(x,p)$ by homodyning on $\vert
\sqrt{N/2}\; x_\alpha\rangle\vert
\sqrt{N/2}\; p_\alpha\rangle $, $\hat{O}_\beta$ a known
semi-definite Hermitian operator , and $\beta=(x+ip)/\sqrt{2}$.
Optimization of this fidelity with respect to the reconstructed
state boils down to finding the largest eigenvalue of this
operator $\hat{O}_\beta$. We can calculate this value analytically
\cite{Braunstein}, from which we obtain a maximum fidelity of
\begin{equation}\label{F*max}
 F_{BS}^{*N}\leq\frac{2N+\lambda}{2N+\lambda+1}
\end{equation}
Equality is achieved by reconstruction of the corresponding
eigenvector $\vert
f_\beta\rangle=\vert\frac{2\sqrt{N}}{2N+\lambda}\beta\rangle $.
Clearly, (\ref{F*max}) is larger than (\ref{Fmax}) for any $N$,
which proves that $F<F^*$. $\Box$

\indent Interestingly, $F_{BS}^{*N}=F^{2N}$, so this nonlocal strategy 
on two states $\vert\alpha\rangle\vert\alpha^*\rangle$ performs as well as
the optimal strategy on four copies of the coherent state $\vert \alpha
\rangle$. This correspondence has a simple explanation. Consider
the input state $\vert \alpha \rangle^{\otimes 4}$. We can send
two of these modes on a beam splitter, and the other two 
on a second beam splitter. This will output the vacuum and two copies
of a coherent state with twice the intensity, 
i.e., $\vert \alpha \rangle^{\otimes
4}\rightarrow \vert \sqrt{2}\alpha \rangle^{\otimes 2}$. We know
from the previous section that the optimal strategy on two copies
of a coherent state consists of independent heterodyning, or
equivalently, measuring the $x$ quadrature on one of the mode, and
$p$ on the other. Indeed, homodyning of the $x$ quadrature on $
\vert \sqrt{2}\alpha \rangle$ or $\vert x_\alpha \rangle$ gives
identical statistics (and the same for $p$), hence the
corresponding fidelities are equal.


\begin{figure}[t] \centering \includegraphics[width=7.5cm]{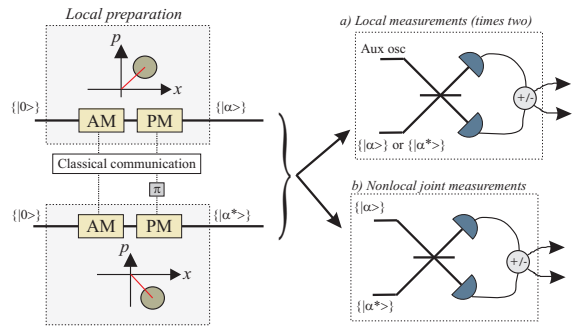} \caption{\it Schematic of the experimental setup. The diagrams show the phase space contours of $|\alpha\rangle$ (upper diagram) and $|\alpha^*\rangle$ (lower diagram). The states are measured using  a) a local strategy and b) a non-local strategy. AM: Amplitude modulator, PM: Phase modulator.}
\label{setup}  \end{figure}


Now, let us proceed with the experimental demonstration 
of the nonlocal feature exhibited by $\{\ket{\alpha}\ket{\alpha^*}\}$. The laser used in our experiment is a monolithic ND:YAG laser producing a field at 1064~nm, which is split into two parts and subsequently directed into the coherent state preparation stage (see Fig.~\ref{setup}). To ensure that the information is encoded as pure coherent states, the states are assumed to be residing at a radio frequency sideband defined within a certain bandwidth of the laser beam. In addition to the high degree of purity, the sideband encoding also holds the advantage of allowing for easy low-voltage control of the coherent amplitudes via simple electro-optic modulators operating at the sideband frequency \cite{exp}. Note therefore that the two beams are bright although the particular sidebands in question are vacuum states, $\{|0\rangle\}$, before the encoding. The production of the two phase conjugate coherent states, $|\alpha\rangle$ and $|\alpha ^*\rangle$, are then performed by displacing the vacuum sidebands using an amplitude modulator (AM) and a phase modulator (PM) in each arm as shown in Fig.~\ref{setup}. 
The two states are prepared by using the same signal generator, that is by communicating classically correlated information between the two preparation stations. The relative phase shift of $\pi$ between the phase quadratures was established by adjusting the cable lengths appropriately.

\begin{figure}[t] \centering \includegraphics[width=8cm]{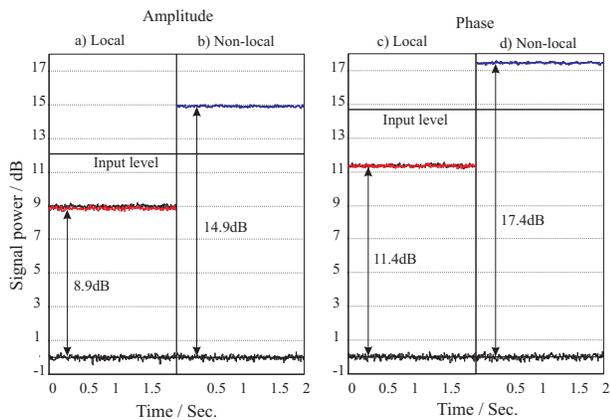} \caption{\it Spectral power densities of the local and nonlocal strategies. All spectra are normalized to the quantum noise level. The resolution bandwidth is 100kHz and the video bandwidth is 30Hz.}\label{results}  \end{figure}

First, we characterize the prepared states by measuring the two copies individually, by successive use of a heterodyne detector yielding information about the amplitude and phase quadratures, simultaneously. The coherent state is combined with a phase stabilized auxiliary beam at a 50:50 beam splitter with a $\pi/2$ relative phase shift and balanced intensities. They interfere with a contrast of ~99\% and the two output beams are detected with high quantum efficiency (95\%) photodiodes. Subsequently the photocurrents are subtracted and added which provides information about the phase and amplitude quadratures, respectively. Finally the spectral densities of the quadratures are recorded on a spectrum analyzer. Using the fact that the heterodyne detector projects the signal under investigation onto a vacuum state, we easily infer the spectral densities of the prepared copies. Furthermore the measurements have also been corrected to account for the detection losses and electronic dark noise in order to avoid an erroneously underestimation. The inferred results for the spectral densities are shown by the solid horizontal lines in Fig.~\ref{results}. 

These measurements for characterization of the prepared copies are in fact identical to the measurements associated with an optimal local estimation strategy. However, in contrast to the characterization, for the estimation of unknown coherent states we are not correcting for detector losses and electronic dark noise. The individual spectral densities for local measurements of $|\alpha\rangle$ and $|\alpha *\rangle$ are shown in column a) and c) of Fig.~\ref{results}. From these measurements we find the added noise to be $\Delta _x=\Delta _p=1.12\pm 0.04$ for the amplitude and phase quadratures. Assuming a flat distribution of coherent states, the fidelity is given by
\begin{equation}
F=\frac{2}{\sqrt{(2+\Delta_x)(2+\Delta_p)}}
\label{fidexp}
\end{equation}
and calculated to $F_L=64.0\pm1\%$. This is close to the theoretical maximum of 2/3.\\

We now discuss the experimental realization of the optimal nonlocal measurement of the phase conjugate copies. As mentioned above, the optimal nonlocal estimation strategy is to combine the two copies at a 50:50 beam splitter and subsequently measure the amplitude quadrature in one output and the phase quadrature in the other output port of the beam splitter. Such a strategy measures the combinations $\hat x_1+\hat x_2$ and $\hat p_1-\hat p_2$ where the indices refer to the two input modes. This combination can, however, be accessed using an experimentally simpler approach since the information is encoded onto sidebands of two equally intense bright beams (with the power 60~$\mu W$). The two classically correlated copies are carefully mode-matched ($\sim$99\%) at a 50:50 beam splitter and actively locked to have balanced intensities at the outputs of the beam splitter. Directly measuring the two outputs yield the quadrature combinations $\hat i_1=(\hat x_1+\hat x_2+\hat p_1+\hat p_2)/2$ and $\hat i_2=(\hat x_1+\hat x_2-\hat p_1-\hat p_2)/2$, and by adding and subtracting these two contributions we obtain the required combinations $\hat x_1+\hat x_2$ and $\hat p_1-\hat p_2$. The spectral densities of these measurements are shown in column b) and d) of Fig.~\ref{results}. 

The upper traces in Fig.~\ref{results} correspond to the coherent amplitudes of the input states and of the joint estimates, whereas the lower traces are the powers associated with the noise levels, all of which are at the shot noise level. The signal-to-noise ratio of the estimate is clearly larger than that of the prepared states; the coherent amplitudes of the amplitude and phase quadratures are increased by 3.0 dB and 2.9 dB, respectively, which effectively correspond to noise equivalent power of $\Delta_x=0.51\pm0.02$ and $\Delta_p=0.52\pm0.02$ shot noise units. Using Eq.~(\ref{fidexp}), the fidelity is calculated to be $F^*=79.5\pm0.7\%$, thus clearly surpassing the classical local fidelity of $2/3$ and close to the theoretical value 4/5.\\

In summary, we have predicted the existence of nonlocal effects on
an ensemble of classically correlated coherent states. We have tested 
this prediction experimentally using a pair of sideband-encoded 
coherent states produced by modulating a CW laser beam. We have unfortunately
not been able to prove the optimality of the joint measurement 
of $\ket{\alpha}\ket{\alpha^*}$ giving $F^*$, because the technique 
we used to prove the optimality of $F$ happened to be hard to adapt.
However, we conjecture that it is the case; this is a topic
for further investigation.

We thank Gerd Leuchs for useful discussions,
and acknowledge financial support from the EU under project
COVAQIAL (FP6-511004). 
AA and MN thank the Spanish MEC, under a ``Ram\'on y Cajal" grant,
and Fundaci\'on Ram\'on Areces for financial support.
NJC, RG-P and JN thank the Belgian government for
financial support under grant IUAP V-18. 
RG-P and JN acknowledge support from the Belgian foundation FRIA.


\begin{thebibliography}{99}
\bibitem{fn} Here, nonlocality is understood in a restricted sense,
not as the incompatibility with any local-hidden-variable model.
\end{thebibliography}
\end{document}